\documentclass[10pt]{article}
\usepackage{amssymb}
\usepackage{setspace}
\usepackage{fancyhdr}
\usepackage{amsfonts,amssymb,latexsym,amsmath,amsthm}
\newtheorem{theorem}{Theorem}[section]
\newtheorem{example}[theorem]{Example}
\newtheorem{lemma}[theorem]{Lemma}
\newtheorem{corollary}[theorem]{Corollary}
\newtheorem{definition}[theorem]{Definition}
\newtheorem{remark}[theorem]{Remark}
\makeatletter\def\blfootnote{\xdef\@thefnmark{}\@footnotetext}\makeatother

\def\span{\mathrm{span}}
\def\bbC{\mathbb{C}}
\def\bbH{\mathbb{H}}
\def\bbCP{\mathbb{CP}}
\def\tr{\mathrm{tr\,}}
\def\cG{\mathcal{G}}
\def\cF{\mathcal{F}}
\def\cI{\mathcal{I}}
\def\cU{\mathcal{U}}
\def\sT{\sf{T}}
\def\Symk{\mathrm{Sym}^k}
\def\Hone{H_{(0,1)}}
\def\H0k{H_{(0,k)}}
\begin{document}

\title{Geometry of the Welch Bounds}


\author{S. Datta\dag, S Howard\ddag, and D. Cochran \S}
%
%
\date{}
\maketitle
\doublespace
\blfootnote{\dag Department of Mathematics, University of Idaho, Moscow ID 83844-1103, USA; \textbf{sdatta@uidaho.edu}}
\blfootnote{\ddag Defense Science and Technology Organisation, PO Box 1500, Edinburgh 5111 SA, Australia; \textbf{Stephen.Howard@dsto.gov.au}}
\blfootnote{\S School of Mathematical and Statistical Sciences, Arizona State University, Tempe AZ 85287-1804, USA; \textbf{cochran@asu.edu}}
\begin{abstract}
A geometric perspective involving Grammian and frame operators is used to derive the entire family of Welch bounds. This perspective
unifies a number of observations that have been made regarding tightness of the bounds and their
connections to symmetric $k$-tensors, tight frames, homogeneous polynomials, and $t$-designs. In particular, a connection has been drawn between sampling of homogeneous polynomials and frames of symmetric $k$-tensors. It is also shown that tightness of the bounds requires tight frames. The lack of tight frames of symmetric $k$-tensors in many cases, however, leads to consideration of sets that come as close as possible to attaining the bounds. The geometric derivation is then extended in the setting of generalized or continuous frames. The Welch bounds for finite sets and countably infinite sets become special cases of this general setting.
\end{abstract}




\section{Introduction} \label{sec:intro}
\subsection{Background}\label{subsec:background}
In a brief but important 1974 paper \cite{W1}, L.~R.~Welch considered the situation of unit vectors
$\{x_1,\ldots,x_m\}$ in $\bbC^n$ with $m>n$. For the maximal cross correlation $c_{\rm{max}}=\max_{i\neq j} |\left< x_i,x_j\right>|$ among the vectors, he developed a family of lower bounds
on $c_{\rm{max}}^{2k},$ parameterized by $k \geq 1,$ given by
\begin{equation}
\label{eq:Welch_weak}
c_{\mathrm{max}}^{2k} \geqslant
\frac{1}{m-1}\left[\frac{m}{\binom{n+k-1}{k}} - 1\right].
\end{equation}
 He described the implications of these bounds in the design of sequences having desirable
correlation properties for multichannel communications applications.  In the years following
their original derivation, the Welch bounds became a standard tool in waveform design for both
communications and radar.
Welch obtained inequality (\ref{eq:Welch_weak})
as a corollary to a more fundamental one:
\begin{equation}
\label{eq:Welch_fundamental}
\sum_{i=1}^m \sum_{j=1}^m |\left< x_i,x_j\right>|^{2k} \geqslant
\frac{m^2}{\binom{n+k-1}{k}}.
\end{equation}
Indeed, most recent work on this topic
recognizes (\ref{eq:Welch_fundamental}) as Welch's main result and refers to
these inequalities as the Welch bounds.

Some variations on this basic result have been noted.  For example, relaxing the
unit-norm assumption \cite{SW1} to allow the $x_i$ to be any non-zero vectors yields
\[
\frac{\displaystyle\sum_{i=1}^m \sum_{j=1}^m |\left<x_i,x_j\right>|^{2k}}
{\left(\displaystyle \sum_{i=1}^m ||x_i||^{2k}\right)^2} \geqslant
\frac{1}{\binom{n+k-1}{k}}.
\]
In any case, the bounds given in (\ref{eq:Welch_fundamental}) are at the heart of the
subject and will be the focus of attention here.

Although Welch's original derivation was analytical, several subsequent authors have noted that the
Welch bounds have a geometric character. Geometric derivations of the first Welch bound, i.e., for $k=1,$ were published
in 2003 by Strohmer and Heath \cite{SH1} and Waldron \cite{SW1}. Shapiro gave a similar argument in
unpublished notes a few years earlier \cite{Shapiro}.
In this paper, the geometric perspective is extended
to derive the entire family of Welch bounds.

For $k = 1,$ conditions under which the Welch bound is attained in (\ref{eq:Welch_weak}) or (\ref{eq:Welch_fundamental}) have been studied explicitly by several authors \cite{MM1,DS1, Xia05,Ding07,SW1,SH1}.
In this case, design methods for sets that meet the bound with equality 
have been proposed \cite{MM1,DS1,SH1,Xia05,Ding07,Holmes04}. In this context, the motivation for identifying
such sets has generally involved communications (e.g., CDMA),
though they also have application in waveform design for radar and sonar.

Welch bound equality sets arise in other application contexts, including quantum information processing and coding theory, as well as in purely mathematical settings. In  quantum information theory, symmetric informationally complete positive operator-valued measures (SIC-POVMs) provide a general model for quantum measurement. Relationships between SIC-POVMs, complex projective $t$-designs, mutually unbiased bases, tight frames, and sets attaining the Welch bounds with equality has been noted in many places \cite{KR1, Ren04, RoyScott07}.
The treatment of Welch bounds for $k\geq1$ in (\ref{eq:Welch_fundamental}) and sets that satisfy them with equality, from the perspective of mutually unbiased bases and $t$-designs, is found in \cite{KR1, Ren04, RoyScott07}. Related results on complex projective $t$-designs, as seen from a more mathematical perspective, is given in other places, for example \cite{Hoggar82}.

This paper gives \emph{frame} conditions for equality in the
Welch bounds in both (\ref{eq:Welch_weak}) and (\ref{eq:Welch_fundamental}), for all $k \geq 1,$ see Section \ref{sec:tight}, and also comments on conditions under which these bounds are relevant. However, constructing such sets for $k >1$ is challenging. It is known, from results, including some in the literature mentioned above, that such sets do not exist in many cases. 
Here the idea of frame potential energy \cite{BF03} is used as the basis for the notion of ``Approximate Welch Bound Equality'' sets.

The existence of isometries between certain spaces of homogeneous
polynomials and symmetric tensors is well established in published work \cite{RS1}.
It is shown here that there is a connection between reconstruction of a homogeneous polynomial from its samples and tight frames of symmetric tensors.

Strohmer and Heath
\cite{SH1} developed Welch-like bounds in infinite-dimensional settings, whereas this paper gives new
results for infinite collections of vectors that form a frame for a finite-dimensional
space. This involves formulation of the Welch bounds in the setting of generalized frames. The results generated here (Section \ref{sec:frames}) seem related to results appearing elsewhere in the literature, notably \cite{RoyScott07}.
\subsection{Outline}\label{subsec:outline}
The foundation of the relationship between the Welch bounds and symmetric $k$-tensors is
elucidated in the derivation of the Welch bounds given in Section \ref{sec:bounds}. It is shown that the derivation can be done using either the Grammian or the frame operator. Section \ref{sec:tight} contains discussion on the construction and existence of Welch Bound Equality (WBE) sets, Maximal Welch Bound Equality (MWBE) sets, and Approximate Welch Bound Equality (AWBE) sets. Section \ref{sec:polynomials}
addresses the connection to homogeneous polynomials and gives a sampling result for homogeneous polynomials.
An extension to generalized frames, which
subsumes both the finite and countably infinite frame cases, is presented in Section \ref{sec:frames}. The section
concludes with some remarks relating tight generalized frames to Haar measures and linking homogeneous
polynomials and $t$-designs.

\subsection{Notation and Terminology}\label{subsec:Notation}
For $x=[x^{(1)}\;\cdots\; x^{(n)}]^{\sT}$ and $y=[y^{(1)}\;\cdots\; y^{(n)}]^{\sT}$ in $\bbC^n$, their inner
product will be denoted by
\[
\left< x,y\right> = \sum_{j=1}^n \overline{x^{(j)}} y^{(j)}
\]
where the bar denotes complex conjugate; i.e., the inner product is conjugate linear in its
first argument and linear in its second argument.  The corresponding convention will be used for inner
products in other complex Hilbert spaces.  Given a finite frame  $\Phi=\{x_1,...,x_m\}$ for an $n$-dimensional
complex vector space $V$, the function $F:V\rightarrow \ell_2(\{1,\ldots,m\})=\bbC^m$ given by
$$F(w)=[\left< x_1,w\right> \;\ldots\; \left< x_m,w\right>]^{\sT}$$ will be called the {\em Bessel map}
associated with $\Phi$, while $\cF=F^* F:V\rightarrow V$ (i.e., the composition of the adjoint of $F$ with $F$)
will be called the {\em frame operator}
associated with $\Phi$.  This terminology will be carried over to the
setting of generalized frames in Section \ref{sec:frames}. For the fundamentals on frame theory the reader is referred to \cite{Chr03, Dau92}.

The space of symmetric $k$-tensors (\cite{RS1}, \cite{Dodson}) associated with a vector space $V$ is denoted by $\Symk(V).$ $\Symk(V)$ is spanned by the tensor powers
$v^{\otimes k}$ where $v\in V$. If $V$ has dimension $n$ then
\[
\dim\ \Symk(V)= \binom{n+k-1}{k} .
\]
$\Symk(V)$ has a natural inner product with the
property
\begin{equation}
\label{eq:symkip}
\left<v^{\otimes k}, w^{\otimes k}\right>_{\Symk(V)} = \left<v,w\right>^k_V  \ .
\end{equation}
The identity map from the vector space $V$ to itself is denoted by $\cI_V.$
\section{The Welch bounds}
\label{sec:bounds}

\subsection{The first Welch bound}

The bound in (\ref{eq:Welch_fundamental}) with $k=1$ has received, by far, the most
attention in the literature.  As noted in Section \ref{subsec:background}, geometric proofs of this particular
bound have appeared in published work and were known as early as 1998 \cite{DL1}.
The
``first Welch bound'', i.e., for $k=1$ in (\ref{eq:Welch_fundamental}), is derived in this section.  This derivation
introduces the essential geometric foundations for obtaining the general case, which is
carried out in the following section.

The following lemma will be key in deriving the Welch bounds.
\begin{lemma}\label{keylemma}
Let $W$ be a finite dimensional subspace of a complex Hilbert space $\bbH$ and
let $T:\bbH \rightarrow \bbH$ be a positive semidefinite linear operator whose range is $W.$\footnote{Most
past work on Welch bounds is set explicitly in $\bbC^n$. It is useful in what follows
to take the slightly more abstract perspective set forth here.}
Denote $n=\dim W$ and let $\lambda_1,\ldots,\lambda_n$ be the non-zero eigenvalues
of $T$.  Then the Hilbert-Schmidt (Frobenius) norm of $T$ satisfies
\begin{equation}
\label{eq:key_inequality}
||T||^2 \geqslant \frac{1}{n}\left(\sum_{i=1}^n |\lambda_i|\right)^2 = \frac{|\tr\ T|^2}{\dim W} \ .
\end{equation}
Equality holds  if and only if all the eigenvalues are equal to each other in which case
$$T = \frac{|\tr\ T|^2}{\dim W} \cI_W.$$
\end{lemma}
The above lemma follows by the Cauchy-Schwarz inequality and the fact that the Frobenius norm of $T$ satisfies
\[
||T||^2 = \sum_{i=1}^n |\lambda_i|^2.
\]
\begin{theorem}[First Welch Bound]\label{FirstWelchBound}
Suppose that $X = \{x_1,\ldots,x_m\}$ is a set of unit vectors in $\bbH$ that span a subspace $V$ of dimension $n$ with $m > n.$ Then
\begin{equation}
\label{eq:Welch_fundamental1}
\sum_{i=1}^m \sum_{j=1}^m |\left< x_i,x_j\right>|^{2} \geqslant
\frac{m^2}{n} \ .
\end{equation}
\end{theorem}
\begin{proof}
Let $F$ be the  Bessel map on $V$ associated with $X.$
Then the Grammian $\cG=FF^*$ is an operator of rank $n$ on $\bbC^m$ whose Frobenius norm
is
\[
||\cG|| = \left(\sum_{i=1}^m \sum_{j=1}^m |\left< x_i,x_j\right>|^{2}\right)^{\frac{1}{2}}
\]
and whose trace is $m$.  Further, the rank of $\cG$ is exactly $n$, so it operates non-trivially
on a subspace $W\subset\bbC^m$ of that dimension.  Thus applying (\ref{eq:key_inequality}) of Lemma \ref{keylemma} to $\cG$ yields
the Welch bound (\ref{eq:Welch_fundamental1}).

A ``dual'' argument is obtained by considering the frame operator $\cF=F^* F:\bbH\rightarrow\bbH$.
The non-zero eigenvalues of $\cF$ are identical to those of $\cG$, so its trace and rank are also equal
to those of $\cG$.  So (\ref{eq:key_inequality}) of Lemma \ref{keylemma} applied to $\cF$ also yields the result.
\end{proof}
%
%

\subsection{Higher-order Welch bounds}\label{High-Order-Welch}

Alternatives to Welch's original analytical derivation of the bounds (\ref{eq:Welch_fundamental})
for $k>1$ do not seem to appear in published literature.  In fact, these cases also follow from
(\ref{eq:key_inequality}) by considering either $k$-fold Hadamard products \cite{HJ1} or tensor products.\footnote{The authors have recently become aware of a manuscript by S. Waldron, in preparation for publication as a book chapter \cite{SWaldron}, that presents a similar perspective to the one set forth in this section.}

\begin{theorem}[Higher Order Welch Bounds]\label{HigherWelchBounds}
Suppose that $X = \{x_1,\ldots,x_m\}$ is a set of unit vectors in $\bbH$ that span a subspace $V$ of dimension $n.$ Then for all integers $k \geqslant 1$
\begin{equation}
\label{eq:Welch_fundamental2}
\sum_{i=1}^m \sum_{j=1}^m |\left< x_i,x_j\right>|^{2k} \geqslant
\frac{m^2}{\binom{n+k-1}{k}} \ .
\end{equation}
\end{theorem}
\begin{proof}
(i) The left-hand side of (\ref{eq:Welch_fundamental2})
is the Hilbert-Schmidt norm of the $k$-fold Hadamard product \cite{HJ1}
$\cG^{\circ k}$
of the Grammian $\cG$ associated with $X.$ From work done in \cite{PW1},
the rank of $\cG^{\circ k}$ is at most $\binom{n+k-1}{k}.$ \footnote{In fact, it has been shown in \cite{PW1} that, almost always, $\textrm{rank}(\cG^{\circ k}) = \min(\binom{n+k-1}{k}, m).$}
The Schur product theorem \cite{HJ1,PW1} ensures that $\cG^{\circ k}$ is positive semidefinite.
Since $\tr\ \cG^{\circ k} = \sum_{i=1}^m ||x_i||^{2k} = m$, (\ref{eq:key_inequality}) gives
\[
||\cG^{\circ k}||^2 = \sum_{i=1}^m \sum_{j=1}^m |\left<x_i, x_j\right>|^{2k}
\geqslant \frac{m^2}{\binom{n+k-1}{k}} \ .
\]

(ii)
Alternatively, consider the space $\Symk(V)$ where $V$ is the $n$-dimensional
span of $X.$ This space has dimension $\binom{n+k-1}{k}$
and the set $X^{(k)}=\{x_1^{\otimes k},\ldots,x_m^{\otimes k}\}$ is a frame for a subspace of $\Symk(V)$.
Denoting the frame operator associated with $X^{(k)}$ by $\cF^{(k)}$,
note that
\[
\tr\ \cF^{(k)} = \sum_{i=1}^m \left< x_i^{\otimes k},x_i^{\otimes k}  \right>
= \sum_{i=1}^m \left< x_i,x_i \right>^k = m \ .
\]
Thus applying inequality (\ref{eq:key_inequality}) of Lemma \ref{keylemma} to  $ \cF^{(k)}$ gives
\[
||\cF^{(k)}||^2 = \sum_{i=1}^m \sum_{j=1}^m |\left<x_i, x_j\right>|^{2k}
\geqslant \frac{m^2}{\binom{n+k-1}{k}}
\]
as desired.
\end{proof}
In the above derivation, the binomial coefficient in the denominator of the Welch bounds has an
explicit geometric interpretation as the dimension of the subspace on which $\cG^{\circ k}$
operates non-trivially.

As already pointed out in Section \ref{subsec:background}, the Welch bounds given by (\ref{eq:Welch_weak}) can be obtained as a corollary to Theorem \ref{HigherWelchBounds}.
\begin{corollary}\label{cor:Welch}
Suppose that $\{x_1,\ldots,x_m\}$ are unit vectors in $\bbH$ that span a subspace $V$ of dimension $n.$ Let $c_{\rm{max}}=\max_{i\neq j} |\left< x_i,x_j\right>|.$ Then for all integers $k \geqslant 1$
\begin{equation*}
c_{\mathrm{max}}^{2k} \geqslant
\frac{1}{m-1}\left[\frac{m}{\binom{n+k-1}{k}} - 1\right].
\end{equation*}
\end{corollary}
\begin{proof}
Due to Theorem \ref{HigherWelchBounds}, (\ref{eq:Welch_fundamental2}) holds and is equivalent
to
\[
\sum_{i\neq j} |\left< x_i,x_j\right>|^{2k} \geqslant
\frac{m^2}{\binom{n+k-1}{k}} - m .
\]
Because the $m(m-1)$ terms in the sum on the left are all non-negative, their maximum must be
at least as large as their average and the result follows.
\end{proof}
\section{Tightness of the Welch bounds}
\label{sec:tight}
As noted in Section \ref{subsec:background}, several authors have investigated
conditions under which the Welch bound with $k=1$ is satisfied with equality.
A condition for all $k\geqslant 1$ is given below, followed by some discussion
about when these higher-order Welch bounds are meaningful.
However, sets that attain the bounds are hard to construct. This naturally leads to the notion of \emph{Approximate Welch Bound Equality} (AWBE) sets, which is also addressed in this section.

As prevalent in the literature \cite{Ding07, Xia05, KR1, DS1, MM1, SH1, SW1}, a set $X$ that meets inequality (\ref{eq:Welch_fundamental}) with
equality is known as a {\em Welch Bound Equality (WBE)} set. If $X$ meets
inequality (\ref{eq:Welch_weak}) with equality, it is called a {\em Maximal Welch
Bound Equality (MWBE)} set \cite{Ding07, Xia05, DS1, SH1}.
%
\subsection{Conditions for Equality}\label{CondforEq}

Conditions for equality to hold in the Cauchy-Schwarz inequality
imply that equality holds in (\ref{eq:key_inequality}) of Lemma \ref{keylemma} if and only if all the
eigenvalues of $T$ are equal. This is used in the following.
\begin{theorem}[Equality in the Welch Bounds]\label{EqualityWelch}
Given a set of unit vectors $X = \{x_1, \ldots, x_m\}$ in $\mathbb{H}$ that span an $n$-dimensional subspace $V,$ let $X^{(k)} = \{x_1^{\otimes k}, \ldots, x_m^{\otimes k}\}.$
Then for integer $k \geq 1,$ \\
(i) $X^{(k)}$ is a WBE set if and only if $X^{(k)}$ is a tight frame for the space $\Symk(V).$
\\
(ii) $X^{(k)}$ is an MWBE set if and only if  $X^{(k)}$ is an equiangular tight frame for the space $\Symk(V).$\\
In both cases the frame bound is $\frac{m}{\binom{n + k -1}{k}}.$
\end{theorem}
\begin{proof} (i)
When $k=1,$ having a WBE set is equivalent to having equality in (\ref{eq:Welch_fundamental1}) of Theorem \ref{FirstWelchBound}. Due to Lemma \ref{keylemma}, this means that that all the non-zero
eigenvalues of the Grammian $\cG$ (and of the frame operator $\cF$) associated with $X$ must be equal to $m/n$.
This holds if and only if $X$ is a tight frame for $V,$ in which case $\cF=\frac{m}{n}\cI_V.$

When $k\geqslant 1$, having a WBE set is equivalent to having equality in (\ref{eq:Welch_fundamental2}) of Theorem \ref{HigherWelchBounds}. The $m\times m$ Gram matrix associated with
the set
$X^{(k)}$ is
\begin{equation*}
\cG_{X^{(k)}} =
 \left[\begin{array}{ccc} \left<x_1^{\otimes k}, x_1^{\otimes
k}\right> & \cdots & \left<x_1^{\otimes k}, x_m^{\otimes k}\right> \\
\vdots & \ddots & \vdots \\
\left<x_m^{\otimes k}, x_1^{\otimes k}\right> & \cdots & \left<x_m^{\otimes k},x_m^{\otimes k}\right>
\end{array} \right]
=\left[\begin{array}{ccc} \left<x_1, x_1\right>^k & \cdots & \left<x_1, x_m\right>^k \\
\vdots  & \ddots & \vdots \\
\left<x_m, x_1\right>^k & \cdots & \left<x_m, x_m\right>^k
\end{array}
\right] 
\end{equation*}
which is the same as the $k$-fold Hadamard product of $\cG,$ i.e, $\cG^{\circ k}.$
From Lemma \ref{keylemma}, equality holds if and only if all the
non-zero eigenvalues of $\cG_{X^{(k)}}$ are equal to
$\frac{m}{\binom{n+k-1}{k}}.$ This is the same as the eigenvalues of the frame operator $\mathcal{F}^{(k)}.$ The set
$X^{(k)}$
is therefore a WBE set if and only if it is a tight frame for
$\Symk(V)$ with frame
operator
\[
\mathcal{F}^{(k)}= \frac{m}{\binom{n+k-1}{k}}\cI_{\Symk(V)} \ .
\]
(ii)
Using the fact that the maximum in a set of non-negative numbers is greater than or equal to the average and the result of Theorem \ref{HigherWelchBounds}, one gets
\[
\max_{i\neq j} |\left< x_i,x_j \right>|^{2k} \geqslant \frac{1}{m(m-1)}\sum_{i\neq j} |\left< x_i,x_j \right>|^{2k} \geqslant \frac{1}{m-1}\left[\frac{m}{\binom{n+k-1}{k}} - 1\right] \ .
\]
Thus, to be a MWBE set, equality must hold in both the inequalities.
In part (i) it has been established that equality holds in the second inequality if and only if $X^{(k)}$ is a tight frame for  $\Symk(V).$
Equality holds in the first inequality if and only if
$|\left< x_i,x_j \right>|$ is constant for all $i\neq j,$ i.e., if and only if the vectors in $X$ are \emph{equiangular} in $\bbH.$ Due to the inner product property (\ref{eq:symkip}), this means that $X^{(k)}$ must be an equiangular set in $\Symk(V),$  when $k \geq 1.$ The set $X^{(k)}$
is therefore a MWBE set if and only if it is an equiangular tight frame for
$\Symk(V).$ The frame bound $\frac{m}{\binom{n+k-1}{k}}$ comes from part (i).

\end{proof}

\subsection{Non-triviality of the bounds}


A necessary condition for (\ref{eq:Welch_fundamental}) to not be vacuous is
that
\[
m > \binom{n+k-1}{k} \ .
\]
For a fixed $n,$ this suggests that  $m > \mathcal{O}(n^k)$ as $k\rightarrow\infty$,
thereby implying that for higher values of $k$ one can hope for
meaningful bounds only when $m \gg n$. Similarly, if $k$ is fixed,
useful bounds require $m > \max(n, \binom{n+k-1}{k})$.  This
implies that $m > \mathcal{O}(k^{n-1})$ as $n\rightarrow\infty$.
In any case, it is evident that the bounds for large $k$ are only
significant when $m\gg n$.

\subsection{Approximate Welch bound equality sequences}\label{apprWBE}

%
Pairs $(m,n)$ for which equiangular tight frames of $m$ vectors in
$\mathbb{C}^n$ can exist along with the required conditions and examples are given in \cite{Sustik07, Bodmann09, Holmes04, Bodmann2010, Tropp05}. Due to Theorem \ref{EqualityWelch}, these are MWBE sets for $k = 1.$ If MWBE sets do not exist for a certain pair $(m,n)$ when $k = 1,$ then MWBE sets of size $m$ also cannot exist for $k>1.$ This is because, by the inner product property (\ref{eq:symkip}), for two pairs $(i, j)$ and $(i', j'),$
$$\left|\langle x_i, x_j \rangle_{\mathbb{C}^n}\right| \neq \left|\langle x_{i'}, x_{j'} \rangle_{\mathbb{C}^n}\right|$$
implies
$$\left|\langle x_i^{\otimes k}, x_j^{\otimes k} \rangle_{\Symk(\bbC^n)}\right| \neq \left|\langle x_{i'}^{\otimes k}, x_{j'}^{\otimes k} \rangle_{\Symk(\bbC^n)}\right|.$$

There are not many values of $k$ for which MWBE sets can be constructed. The maximum number of equiangular lines in $\mathbb{C}^n$ is $n^2$ \cite{DGS75, LintSeidel66}. Due to (\ref{eq:symkip}), equiangular lines in $\Symk(\bbC^n)$ are also equiangular lines in $\mathbb{C}^n$ and so a necessary condition for the existence of MWBE sets is 
\begin{equation}\label{MWBEexistence}
\binom{n+k-1}{k} \leq n^2.
\end{equation}
For a fixed dimension $n$, there are not many values of $k$ that satisfy (\ref{MWBEexistence}); in fact, for $n\geqslant 3$ there is no $k >2$ for which MWBE sets can exist.

However, for $k\leq2$ and $m = n^2,$ MWBE sets are the same as symmetric informationally complete positive operator-valued measures (SIC-POVMs), which have been studied extensively in connection to quantum measurement \cite{Ren04, KR1, RoyScott07}. SIC-POVMs are hard to construct \cite{KR1, Ren04}; existence of SIC-POVMs in all dimensions $n$ has been conjectured \cite{Zauner99, Ren04, SG10}.
%
%

%
%

In the context of sets that are \emph{not} MWBE sets, for $k=1$, the author in \cite{Waldron09} addresses the construction of equiangular frames for $\mathbb{R}^n$ that are \emph{not} tight and comments on the potential application of these in signal processing and quantum information theory.

When $k \leq 2$, mutually unbiased bases (MUBs) give rise to WBE sets of $n(n+1)$ elements in $\bbC^n$ (Theorem 3, \cite{KR1}) but these are challenging to construct \cite{KR1}. Theorem \ref{EqualityWelch} characterizes WBE sets, for $k \geq 1$, in terms of tight frames for the space of symmetric $k$-tensors. These sets are also equivalent to complex projective $k$-designs (Theorem 2, \cite{Ren04} and Theorem 1, \cite{KR1}). However, complex projective $k$-designs and hence WBE sets (for $k > 2$)
%
are hard to find and known not to exist in many cases \cite{RoyScott07}. Consequently, it seems reasonable to look for sets that are as close as possible to attaining the bound in (\ref{eq:Welch_fundamental}) for a given $k$.
\begin{definition}[Approximate Welch Bound Equality Sets]\label{Def:AWBE}
Let $V$ be an $n$-dimensional subspace of a Hilbert space $\mathbb{H}.$ For $k \geq 1,$ if a set $X = \{x_1,\ldots,x_m\}$ of unit vectors in $V$ minimizes $\sum_{i=1}^m \sum_{j=1}^m |\left<x_i^{\otimes k}, x_j^{\otimes k}\right>|^{2}$ then $X^{(k)} = \{x_1^{\otimes k}, \ldots,$ $x_m^{\otimes k}\}$ is called an \emph{Approximate} Welch Bound Equality (AWBE) set.
\end{definition}
 Definition \ref{Def:AWBE} is inspired by the notion of the frame potential and its connection to tight frames in $\mathbb{C}^n$ \cite{BF03}. The quantity $\sum_{i=1}^m \sum_{j=1}^m |\left<x_i^{\otimes k}, x_j^{\otimes k}\right>|^{2}$ is the frame potential of the set $X^{(k)} = \{x_1^{\otimes k}, \ldots, x_m^{\otimes k}\}$ in $\Symk(\bbC^n).$ When $k=1,$ the minimizer of the frame potential is a tight frame for $\mathbb{C}^n$ (Theorem 7.1, \cite{BF03}) and hence a WBE set.
\begin{example}[Approximate WBE sets]

 \rm
To illustrate numerically the construction of an AWBE set via minimization of the frame potential, consider the case with $k=3$, $n=2$ and $m=7$. To construct an AWBE set with these parameters, start with a frame $X_0 = \{x_1, \ldots, x_7\}$ of seven unit vectors in $\bbC^2$, such as
\[
X_0 = \left\{\left[\begin{array}{c}
1 \\
0
\end{array}\right], \left[\begin{array}{c}
0 \\
1
\end{array}\right],  \left[\begin{array}{c}
\frac{1}{\sqrt{2}} \\
\frac{1}{\sqrt{2}}
\end{array}\right], \left[\begin{array}{c}
-\frac{1}{\sqrt{2}} \\
\frac{1}{\sqrt{2}}
\end{array}\right], \left[\begin{array}{c}
\frac{\sqrt{3}}{2} \\
-\frac{1}{2}
\end{array}\right], \left[\begin{array}{c}
-\frac{\sqrt{3}}{2} \\
-\frac{1}{2}
\end{array}\right], \left[\begin{array}{c}
-\frac{1}{2}\\
\frac{\sqrt{3}}{2}
\end{array}\right]\right\}.
\]
Noting that $\sum_{i=1}^7 \sum_{j=1}^7 |\left<x_i^{\otimes 3}, x_j^{\otimes 3}\right>|^{2} = \sum_{i=1}^7 \sum_{j=1}^7 |\left<x_i, x_j\right>|^{6}$ and using $X_0$ as the initial point, solve
$$\min\sum_{i=1}^7 \sum_{j=1}^7 |\left<x_i, x_j\right>|^{6}$$ subject to the constraint that the vectors $x_i$ in the solution are unit normed. The solution,
obtained by Matlab, is the set
\[
X = \left[
\begin{array}{ccccccc}
 0.99  &  0.14 &   0.56 &  -0.68 &   0.93  & -0.86 &  -0.30 \\
    0.08  &  0.99  &  0.83  &  0.73  & -0.36  & -0.50  &  0.95
 \end{array}
\right].
\]
The ratio of the upper and lower bounds for the frame obtained from $X^{(3)}$ in $\mathrm{Sym}^3(\bbC^2)$ (i.e., the condition number of the frame operator) is $3$. Although not tight, $X^{(3)}$ is as close to tight, in the sense of frame potential energy discussed above, as any frame  of seven unit vectors that are pure tensors in the space $\mathrm{Sym}^3(\bbC^2)$ can be. $X^{(3)}$ is an AWBE set of seven vectors.

It is noteworthy that the set $X$ has lower and upper frame bound equal to $3.5.$ Within the numerical precision of this example, $X$ is thus a tight frame for $\bbC^2$, and the AWBE set $X^{(3)}$ arises as the tensor powers of the elements of a tight frame.  Indeed, this phenomenon has been observed consistently by the authors in numerous numerical experiments of this type.  Tight frames in $V$ minimize $\sum_i \sum_j |\left<x_i, x_j \right>|^2$ while AWBE sets in $\Symk(V)$ minimize
$\sum_i \sum_j |\left<x_i^{\otimes k}, x_j^{\otimes k}\right>|^{2}=\sum_i \sum_j |\left<x_i, x_j\right>|^{2k}$.
\end{example}
\section{Connection to Sampling of Homogeneous Polynomials and Compressed Sensing}
\label{sec:polynomials}

It is well known (see, e.g., \cite{RS1}) that $\H0k$, the linear space of homogeneous polynomials of
total degree $k$ in variables $\bar{z}^{(1)},\ldots,\bar{z}^{(n)}$ is isomorphic to $\Symk(V)$.
This section, in Theorem \ref{thm:sampling} below, points out a connection between the condition that a set
is a frame for $\Symk(V)$ and the reconstructability
of polynomials in $\H0k$ from the values they take at sets of $m$ points in $\bbC^n$.
\begin{theorem}\label{thm:sampling}
Let $X=\{x_1,\ldots,x_m\}$ be a set of vectors in $V.$ For $k \geqslant 1,$ polynomials in $\H0k$ are uniquely
determined from their samples at the points in $X$
if and only if $X^{(k)} = \{x_1^{\otimes k},\ldots,x_m^{\otimes k}\}$ is a frame for $\Symk(V)$.
\end{theorem}
\begin{proof}
(i)
Let $k=1$ and $w\in V=\mathrm{Sym}^1(V).$ Denote by $[w^{(1)}\,\cdots\,w^{(n)}]^{\sT}$ the
coordinates of $w$ in some orthonormal basis for $V$.
 The mapping $p: V \to \Hone$ defined by
 $$p(w)=w^{(1)}\bar{z}^{(1)}+\cdots w^{(n)}\bar{z}^{(n)} = p_w(z^{(1)},\ldots,z^{(n)})$$
is an isomorphism that
takes $w\in V$ to the polynomial $p_w\in\Hone.$

If $X$ is a frame for $V$, the associated Bessel map
$F: V \rightarrow \bbC^m $ is given by
\begin{equation}
\label{eq:frame_poly}
F(w)=\left[ \begin{array}{c}
\left< x_1,w \right> \\
\vdots \\
\left< x_m,w \right> \end{array} \right]
=
\left[ \begin{array}{c}
p_w(x_1^{(1)},\ldots,x_1^{(n)}) \\
\vdots \\
p_w(x_m^{(1)},\ldots,x_m^{(n)}) \end{array} \right] .
\end{equation}
$F(w)$ is a vector of values obtained by evaluating or sampling
$p_w$ at $x_1,\ldots,x_m$.
Define a sampling function $P_X:\Hone\rightarrow\bbC^m$ by
\[
P_X(p)=\left[ \begin{array}{c}
p(x_1^{(1)},\ldots ,x_1^{(n)}) \\
\vdots \\
p(x_m^{(1)},\ldots ,x_m^{(n)}) \end{array} \right].
\]
Note that (\ref{eq:frame_poly}) shows that the Bessel map is given by
$F(w)=P_X(p_w)$.  Because $F$ is invertible, $w$ is \emph{uniquely}
determined by $F(w)$.  Hence any $p_w\in\Hone$ is uniquely determined by its
samples $P_X(p_w)$.

Conversely, if $X$ does not form a frame for $V$, the mapping $F$ defined by (\ref{eq:frame_poly}) has a non-trivial kernel $K$.  In this case,
$P_X(p_w)=P_X(p_{w+u})$ for all $u\in K$.  Therefore, $p_w$ is not uniquely
determined from its samples at $x_1,...,x_m$.

(ii)
For $k>1$, the space of interest is $\Symk(V)$ and the
frame is $X^{(k)}.$ As for $k=1,$
mapping a polynomial
to its coefficient sequence defines an embedding of $\H0k$ in $\Symk(V).$
If $v=w^{\otimes k}\in\Symk(V)$ is a pure tensor power of $w\in V$, then
\begin{equation*}
F^{(k)}(v)  =  \left[ \begin{array}{c}
\left< x_1^{\otimes k}, w^{\otimes k} \right>\\
\vdots \\
\left< x_m^{\otimes k}, w^{\otimes k} \right> \end{array} \right]
 =
\left[ \begin{array}{c}
\left< x_1, w \right>^k\\
\vdots \\
\left< x_m, w \right>^k \end{array} \right]
= \left[ \begin{array}{c}
p_v(x_1) \\
\vdots \\
p_v(x_m) \end{array} \right]
\end{equation*}
where $p_v \in \H0k$
defined by $p_v(z)=\left< z,w\right>^k$.
$\Symk(V)$ is spanned by pure tensor powers of elements in $V$ \cite{RS1}.
Thus, for arbitrary $v\in\Symk(V)$, $F^{(k)}(v)$ is a vector of $m$ samples of a polynomial
in $\H0k$ taken at points
$x_1,...,x_m$.  Thus, a polynomial $p \in \H0k$ is uniquely
determined by the sample set
\[
P_X^{(k)}(p)=\left[ \begin{array}{c}
p(x_1) \\
\vdots \\
p(x_m) \end{array} \right]
\]
if and only if $X^{(k)}$ is a frame for $\Symk(V)$.
\end{proof}
\begin{remark}[\emph{Connection to compressed sensing}] \rm
A signal $x\in \mathbb{C}^N$ is $k$-sparse in a basis $\Psi=\{\psi_j\}_{j=1}^N$ if $x$ is a weighted superposition of at most $k$ elements of $\Psi$. Compressed sensing is broadly concerned with the inverse problem of reconstructing such a signal $x$ from linear measurements
$\{y_\ell=\langle x,\phi_\ell\rangle \ | \ \ell=1,\ldots, n\}$ with $n\ll N$. In the general setting, one has $\Phi x = y$, where $\Phi$ is a $n \times N$ sensing matrix having the measurement vectors $\phi_\ell$ as its rows, $x$ is a length-$N$ signal and $y$ is a length-$n$ measurement.

The standard compressed sensing technique guarantees exact recovery of the signal $x$ with high probability if $\Phi$ satisfies the Restricted Isometry Property (RIP) \cite{Candes_06,CRT06, Don06}. This means that for a fixed $k$, there exists a small number $\delta_k$, such that
\begin{equation*}
 (1-\delta_k)\|x\|^2_{\ell_2}\leq\|\Phi x\|^2_{\ell_2}\leq(1+\delta_k)\|x\|^2_{\ell_2},
\end{equation*}
for any $k$-sparse signal $x$. Denoting any $n \times k$ submatrix of $\Phi$  by $\Phi_T,$ the above is equivalent to saying that all the eigenvalues of $\Phi_T^* \Phi_T$ must lie in $[1 - \delta_k, 1 + \delta_k],$ or, that the rows of $\Phi_T$ form a frame with frame bounds very close to each other and to 1.  From the characterization of WBE sets given in Theorem \ref{EqualityWelch}, this means that if the rows of $\Phi_T$ is a WBE set of $n$ vectors in $\mathbb{C}^k$ then $\Phi$ is a ``good'' sensing matrix.

\end{remark}
%
%
%

\section{Generalized Frames}
\label{sec:frames}


Let $\bbH$ be a complex Hilbert space and $(M,\mu)$ a measure space.
A \emph{generalized frame} in $\bbH$
indexed by $M$ is a family of vectors $\{x_{\alpha} \in \bbH: \alpha \in M\},$ denoted by $(X_M, \mu)$ or just $X_M,$ such that:
\begin{itemize}
\item[(a)] For every $y \in \bbH$, the function
$\tilde{y}: M \to\bbC$ defined by
\[
\tilde{y}(\alpha)=
\left<x_{\alpha}, y \right>_\bbH
\]
is $\mu$-measurable.
\item[(b)]There exist constants $0 < A \leqslant B <\infty$ such that, for every $y \in \bbH$,
\[
A ||y||^2_{\bbH} \leqslant  \int_M |\left<x_{\alpha},y\right>_{\bbH}|^2 d\mu(\alpha)
\leqslant B ||y||^2_{\bbH}
\]
or,
\[
\quad A ||y||^2_{\bbH} \leqslant
||\tilde{y}||^2_{L^2(M,\mu)} \leqslant B ||y||^2_{\bbH} \ .
\]
\end{itemize}
%

The mapping  $F : \bbH \to L^2(M, \mu)$ is
given by $F(y) = \left<x_{\alpha}, y\right>\Big|_{\alpha \in M}$ and
its adjoint is $F^*: L^2(M, \mu) \to \bbH$ with
$F^* (g) = \int_M g(\alpha)x_{\alpha}d\mu(\alpha)$.
The frame operator $\cF:\bbH\to\bbH$ is $\mathcal{F} = F^*F$; i.e., for
$y \in\bbH$
\[
\mathcal{F}(y) = \int_M \left<x_{\alpha}, y\right>x_{\alpha}\;d\mu(\alpha) \ .
\]
The Grammian $\cG:L^2(M,\mu)\to L^2(M,\mu)$ is defined by $\cG = F
F ^*$; i.e.,
\[
(\cG f)(\beta) = \int_M \left<x_{\beta}, x_{\alpha}\right>
f(\alpha)\; d\mu(\alpha) \ .
\]
A good overview of generalized frames is given in
\cite{GK1}.
\subsection{Welch bounds for generalized frames} \label{sec:gen_fr}

With $V$ an $n$-dimensional subspace of $\bbH$, denote by
$S^{n-1}$ the set of unit vectors in $V$. For each $x\in S^{n-1}$,
the mapping $\Pi_x: V \to \span(x)$ given by
\[
\Pi_x(v) = \left<x, v\right>x
\]
is a projector that maps $V$ onto the one-dimensional
subspace spanned by $x.$
Since $\Pi_x = \Pi_{e^{i\theta}x}$ for any $\theta\in [0, 2\pi)$, the collection of
projectors $\Pi_x$ is parameterized by the complex projective space
$\bbCP^{n-1}$.  Given a
normalized measure $\mu$ on $\bbCP^{n-1}$ (i.e., with $\mu(\bbCP^{n-1})=1$),
a generalized frame $X_{\bbCP^{n-1}}$ for $V$ is obtained by selecting
one representative $x\in\bbH$ from each equivalence class corresponding
to a point in $\bbCP^{n-1}$.  The frame operator $\cF_\mu: V \to V$
for this generalized frame is given by
\[
\cF_\mu (v) =
\int_{\bbCP^{n-1}} \Pi_x(v) d\mu(x) =
\int_{\bbCP^{n-1}} \left<x, v\right> x \ d\mu(x) \ .
\]
\begin{theorem}[First Welch Bound]\label{FirstWelchBoundGeneral}
Let $\mu$ be a normalized measure on $\bbCP^{n-1}$ and $X_{\bbCP^{n-1}}$ be a generalized frame for an $n$-dimensional subspace $V$ of a Hilbert space $\bbH.$ Then
\[
\iint_{\bbCP^{n-1}} |\left< x, y \right>|^2 d\mu(x) d\mu(y) \geqslant \frac{1}{n},
\]
with equality if and only if the frame is tight.
\end{theorem}
\begin{proof}
Taking $\{e_1,\ldots,e_n\}$ to be an orthonormal basis of $V$,
the trace of $\cF_\mu$ is given by
\[
\tr\cF_\mu = \sum_{k=1}^n \left<\cF_\mu(e_k), e_k\right> = 1 .
\]
The Hilbert-Schmidt norm of $\cF_\mu$ is
\[
||\cF_\mu||^2_V = \tr\cF_\mu^* \cF_\mu = \iint_{\bbCP^{n-1}} |\left< x, y \right>|^2 d\mu(x) d\mu(y).
\]
Using Lemma \ref{keylemma} in this setting for $k=1$ gives
\[
\iint_{\bbCP^{n-1}} |\left< x, y \right>|^2 d\mu(x) d\mu(y) \geqslant \frac{1}{n} \ .
\]
The bound is achieved if and
only if
\[
\cF_\mu =\frac{1}{n}\cI_V ,
\]
i.e., if and only if the generalized frame is tight.
\end{proof}
For $k\geqslant 1$, higher-order Welch bounds in the
generalized frame setting are obtained by considering
$\Symk(V)$.
The projector $\Pi_{x^{\otimes k}}$ maps $\Symk(V)$ onto the
one-dimensional subspace spanned by the tensor power $x^{\otimes k},$ with $x\in S^{n-1}$.
Direct calculation using (\ref{eq:symkip}) yields
\[
\Pi_{x^{\otimes k}} = \Pi_{x}^{\otimes k},
\]
and, for $v\in V$,
\[
\Pi^{\otimes k}_{x} v^{\otimes k} = \left<x, v \right>^k
x^{\otimes k} \ .
\]
This collection of projectors is parameterized by
$\bbCP^{n-1}$. Corresponding to each $x\in\bbCP^{n-1}$, choosing a
representative unit vector in $V$ yields a collection of unit
vectors
\[
X_{\bbCP^{n-1}}^{(k)} = \{u_x^{\otimes k}|
 \ u_x \in V, x \in\bbCP^{n-1}\} .
\]
Given a normalized measure $\mu$ on $\bbCP^{n-1}$,
$X_{\bbCP^{n-1}}^{(k)}$ becomes a generalized frame for
$\Symk(V)$ with frame operator
$\mathcal{F}_{\mu}^{(k)}:\Symk(V)\rightarrow\Symk(V)$ given by
\[
\mathcal{F}_{\mu}^{(k)} = \int_{\bbCP^{n-1}}
\Pi_{x^{\otimes k}} d\mu(x) \ .
\]
\begin{theorem}[Higher Welch Bounds]\label{HigherWelchBoundGeneral}
Let $\mu$ be a normalized measure on $\bbCP^{n-1}$ and let $X_{\bbCP^{n-1}}$ be a generalized frame for an $n$-dimensional subspace $V$ of a Hilbert space $\bbH.$ Then for all $k \geq 1,$
\begin{equation}\label{eq:Welch_generalized}
\iint_{\bbCP^{n-1}} |\left< x, y\right>|^{2k} d\mu(x) d\mu(y)
\geqslant  \frac{1}{\binom{n+k-1}{k}},
\end{equation}
with equality if and only if
$(X_{\bbCP^{n-1}}^{(k)}, \mu)$ is a
generalized tight frame for $\Symk(V).$
\end{theorem}
\begin{proof}
Noting that $\tr\cF^{(k)}_\mu = 1$, Lemma \ref{keylemma} implies
\begin{equation*}
||\cF^{(k)}_\mu||_{\Symk(V)}^2  =
\iint_{\bbCP^{n-1}} |\left< x, y\right>|^{2k} d\mu(x) d\mu(y) \nonumber \\
 \geqslant  \frac{1}{\binom{n+k-1}{k}}
\end{equation*}
with equality if and only if
\[
\cF_\mu^{(k)} = \frac{1}{\binom{n+k-1}{k}}\cI_{\Symk(V)} ,
\]
i.e., if and only if $(X_{\bbCP^{n-1}}^{(k)}, \mu)$ is a
generalized tight frame for $\Symk(V).$
\end{proof}
\begin{example}\label{ex:finiteframe}
\rm
Let $X=\{x_1,\ldots,x_m\}$ be a set of unit vectors
that is a frame for an $n$-dimensional subspace $V$ of $\mathbb{H}.$ Consider the (normalized) discrete measure
\[
\mu = \frac{1}{m}\sum_{x \in X}\delta_x \ .
\]
Using this measure in Theorem \ref{HigherWelchBoundGeneral}
yields
\[
\frac{1}{m^2}\sum_{x,y \in X}|\left<x,y\right>|^{2k} \geqslant
\frac{1}{\binom{n+k-1}{k}}
\]
which is equivalent to (\ref{eq:Welch_fundamental}).
Equality is obtained if and only if
$X^{(k)} = \{x^{\otimes k} | \ x \in X\}$ is a tight frame for
$\Symk(V),$ i.e., if and only if
\[
\frac{1}{m}\sum_{x\in X} \Pi_{x^{\otimes k}} = \frac{1}{\binom{n+k-1}{k}}\cI_{\Symk(V)} \ .
\]
Thus the generalized frame perspective yields the Welch bound
for finite frames as a special case.
\end{example}
\begin{example}
\rm
If $X = \{x_i\}_{i=1}^{\infty}$
in $\bbCP^{n-1}$ and $\{w_i\}_{i=1}^{\infty}$ is a summable set of positive numbers,
defining a discrete measure by
\[
\mu = \frac{\sum_{i=1}^{\infty} w_i \delta_{x_i}}{\sum_{j=1}^{\infty}w_j}
\]
yields a generalized frame $(X,\mu)$. With this measure in Theorem \ref{HigherWelchBoundGeneral}, one gets
\[
\frac{1}{\left(\sum_{j=1}^{\infty}w_j\right)^2}\sum_{i,\ell}|\left<w_i x_i, w_\ell x_\ell\right>|^{2k}
\geqslant \frac{1}{\binom{n+k-1}{k}} \ .
\]
Thus the generalized frame perspective also produces Welch bounds for countably infinite frames.
\end{example}
%
%

\subsection{Tight generalized frames and Haar measure}

Roy and Scott \cite{RS1} discuss relationships between Haar measure on the $n$-dimensional unitary group $\cU_n$, the unique unitarily invariant
probability measure it induces on $\bbCP^{n-1}$, and unweighted $t$-designs. The formulation in the preceding section enables a frame-theoretic perspective on this circle of ideas.

$\cU_n$ acts transitively on $\bbCP^{n-1}$ and for each $y\in V^{\otimes k}$ define
\[
\Phi^{(k)}_U(y) = U^{\otimes k} y 
\]
where $U\in\cU_n$ \cite{AMP1}.
$\Symk(V)$ is an invariant subspace of $V^{\otimes k}$ under this action.

Equality in (\ref{eq:Welch_generalized}) is attained if and only if $(X_{\bbCP^{n-1}}^{(k)}, \mu)$ is a
generalized tight frame for $\Symk(V)$. It is now shown that this
occurs when $\mu$ is the Haar measure.
\begin{theorem}
Let $\mu$ be the
normalized $\cU_n$-invariant Haar measure on $\bbCP^{n-1}$. Then, for all $k \geqslant 1,$ $(X_{\bbCP^{n-1}}^{(k)}, \mu)$ is a generalized tight frame for $\Symk(V),$ i.e., $$\cF_{\mu}^{(k)} =\frac{1}{\binom{n+k-1}{k}}\cI_{\Symk(V)}.$$
\end{theorem}
\begin{proof}
For any $U\in\cU_n$,
\begin{eqnarray*}
U^{\otimes k} \cF_{\mu}^{(k)} (U^{\otimes k})^*
   & = & \int_{\bbCP^{n-1}} U^{\otimes k}\Pi_x^{\otimes k} (U^{\otimes k})^*\,d\mu(x) \\
   & = & \int_{\bbCP^{n-1}} \Pi_{Ux}^{\otimes k}\,d\mu(x) \\
   & = & \int_{\bbCP^{n-1}} \Pi_x^{\otimes k}\,d\mu(x)
\end{eqnarray*}
where the last equality holds because $\mu$ is $\cU_n$-invariant.  This shows that $\cF_{\mu}^{(k)}$
commutes with all $U^{\otimes k}$.  Since $\Phi_U^{(k)}$ acts on $\Symk(V)$ irreducibly, Schur's lemma implies
$\cF_{\mu}^{(k)}=\lambda\cI_{\Symk(V)}$.  Because $\tr \cF_{\mu}^{(k)}=1$, dimensionality
considerations imply that
\[
\lambda = \frac{1}{\dim\Symk(V)}=\frac{1}{\binom{n+k-1}{k}} \ .
\]
\end{proof}

\subsection{Homogeneous polynomials and $t$-designs}

As in the finite frame case, the generalized frame perspective yields connections to homogeneous
polynomials and, further, to spherical $t$-designs.
Suppose that $(X_{\bbCP^{n-1}}^{(k)},\mu)$ is a generalized tight frame for $\Symk(V)$. Then
\begin{equation}
 \label{eq:met_tight}
\cF_\mu^{(k)} = \int_{\bbCP^{n-1}}
\Pi_{x^{\otimes k}} d\mu(x) = \frac{1}{\binom{n+k-1}{k}}\cI_{\Symk(V)} \ .
\end{equation}
The mapping $F^{(k)} : \Symk(V) \to L^2(\bbCP^{n-1}, \mu )$
is given by $F^{(k)}(w) =  \left< x^{\otimes k}, w \right> \big|_{x\in\bbCP^{n-1}}$ for $w\in\Symk(V)$.
Since the tensor powers $\{v^{\otimes k} : v\in V\}$ span $\Symk(V)$, $\left< x^{\otimes k},w\right>$
can be written as a linear combination of terms of the form $\left< x,w\right>^k$ and
hence is in $H_{(0, k)}$. Denoting this polynomial associated with $w$ by $p_w$ and using (\ref{eq:met_tight})
gives, for any $v,w\in\Symk(V)$,
\begin{eqnarray*}
\left<v, w \right> & = &  \binom{n+k-1}{k} \int_{\bbCP^{n-1}} \left< v, x^{\otimes k}\right> \left< x^{\otimes k}, w \right>\,d\mu(x) \\
& = & \binom{n+k-1}{k} \int_{\bbCP^{n-1}} \overline{p_v(x)} p_w(x)\;d\mu(x) \ .
\end{eqnarray*}

If $\mu$ is the normalized discrete measure discussed in Example \ref{ex:finiteframe} 
the frame operator $\cF_\mu^{(k)}$ can be written as
\[
 \frac{1}{m}\sum_{x\in X} \Pi_{x^{\otimes k}}
 = \frac{1}{\binom{n+k-1}{k}}\cI_{\Symk(V)} \ .
\]
Using this representation of $\cI_{\Symk(V)}$ gives
\[
\left< v,w\right> = \frac{\binom{n+k-1}{k}}{m} \sum_{x\in X}\overline{p_v(x)} p_w(x)
\]
so that
\[
\int_{\bbCP^{n-1}} \overline{p_v(x)} p_w(x)\,d\mu(x) = \frac{1}{m} \sum_{x\in X}\overline{p_v(x)} p_w(x) \ .
\]
This implies that, for any $g\in H_{(k,k)}$, the space of homogeneous polynomials of total
degree $k$ in $x_1, \ldots, x_m$ and total degree $k$ in $\bar{x}_1, \ldots, \bar{x}_m$,
\[
\int_{\bbCP^{n-1}} g(x)\;d\mu(x) = \frac{1}{m} \sum_{x\in X} g(x) \ .
\]
If $X^{(k)}=\{x_1^{\otimes k},\ldots,x_m^{\otimes k}\}$ is a tight frame for $\Symk(V)$ for all $k\leqslant t$,
then
\begin{equation}
\label{eq:tdesign}
\int_{\bbCP^{n-1}} g(x)\;d\mu(x) = \frac{1}{m} \sum_{x\in X} g(x)
\end{equation}
for all $g\in\bigoplus^t_{k=1} H_{(k,k)}$. Equation (\ref{eq:tdesign}) defines
$X^{(k)}=\{x_1^{\otimes k},\ldots,x_m^{\otimes k}\}$ as a complex projective $t$-design \cite{KR1}.

\section{Conclusion}

The classical Welch bounds have been shown to arise from dimensionality considerations in connection
with frame and Grammian operators.
Geometric derivations of the first Welch bound have been given
in previous work.  This paper has extended the geometric perspective to obtain the higher-order Welch bounds,
with the $k^{\mathrm{th}}$ bound for $k\geqslant 1$ arising naturally from observing either the $k$-fold
Hadamard product of the Grammian or the frame operator associated with a frame on a space of
symmetric $k$-tensors.

Welch bounds for generalized frames have been derived and the classical case shown to follow from this
more general result.  The role of tight frames in achieving the Welch bounds with equality has been established in this general setting.  In general, for $k\geq 2,$ due to the difficulty in construction and, in some cases, lack of existence of Welch bound equality sets, it is natural to construct sets that come as close as possible to attaining the bound. This was done here by considering sets that minimize the frame potential energy in the space $\Symk(V)$ of symmetric $k$-tensors, and has led to an open question regarding conditions under which such sets must arise from tight frames of $V$.  Further, specific
connections have been clarified between the circle of ideas entailed in the geometric understanding of the
Welch bounds and related topics involving symmetric tensors, homogeneous polynomials, and $t$-designs. In particular, it has been shown that a homogeneous polynomial of a known degree $k$ can be uniquely reconstructed from its samples, provided the sampling points form a frame for the space of
symmetric $k$-tensors.
\section{Acknowledgements}
\label{Acknowledge}
The authors would like to thank the anonymous reviewers for their insightful comments and valuable suggestions.



\bibliographystyle{plain}
\bibliography{welch_refs}

\end{document}